# PLATE FIN HEAT EXCHANGER MODEL WITH AXIAL CONDUCTION AND VARIABLE PROPERITES

B. J. Hansen, M. J. White, and A. Klebaner

Fermi National Accelerator Laboratory
Batavia, IL, 60510, USA

## ABSTRACT

Future superconducting radio frequency (SRF) cavities, as part of Project X at Fermilab, will be cooled to superfluid helium temperatures by a cryogenic distribution system supplying cold supercritical helium. To reduce vapor fraction during the final Joule-Thomson (J-T) expansion into the superfluid helium cooling bath, counter-flow, plate-fin heat exchangers will be utilized. Due to their compact size and ease of fabrication, plate-fin heat exchangers are an effective option. However, the design of compact and high-effectiveness cryogenic heat exchangers operating at liquid helium temperatures requires consideration of axial heat conduction along the direction of flow, in addition to variable fluid properties. Here we present a numerical model that includes the effects of axial conduction and variable properties for a plate fin heat exchanger. The model is used to guide design decisions on heat exchanger material choice and geometry. In addition, the J-T expansion process is modeled with the heat exchanger to analyze the effect of heat load and cryogenic supply parameters.

**KEYWORDS:** Plate-Fin, Heat Exchanger, Axial Conduction, Numerical Model, Superfluid, Helium, Cryogenic

## INTRODUCTION

As part of Project X, SRF cavities will be cooled to superfluid helium temperatures by a cryogenic distribution system that supplies cold supercritical helium. A counter-flow, plate-fin heat exchanger (PFHE) will be used to further cool the incoming supercritical helium by recuperating some of the sensible heat from the low pressure vapor return of the superfluid helium bath (FIGURE 1).

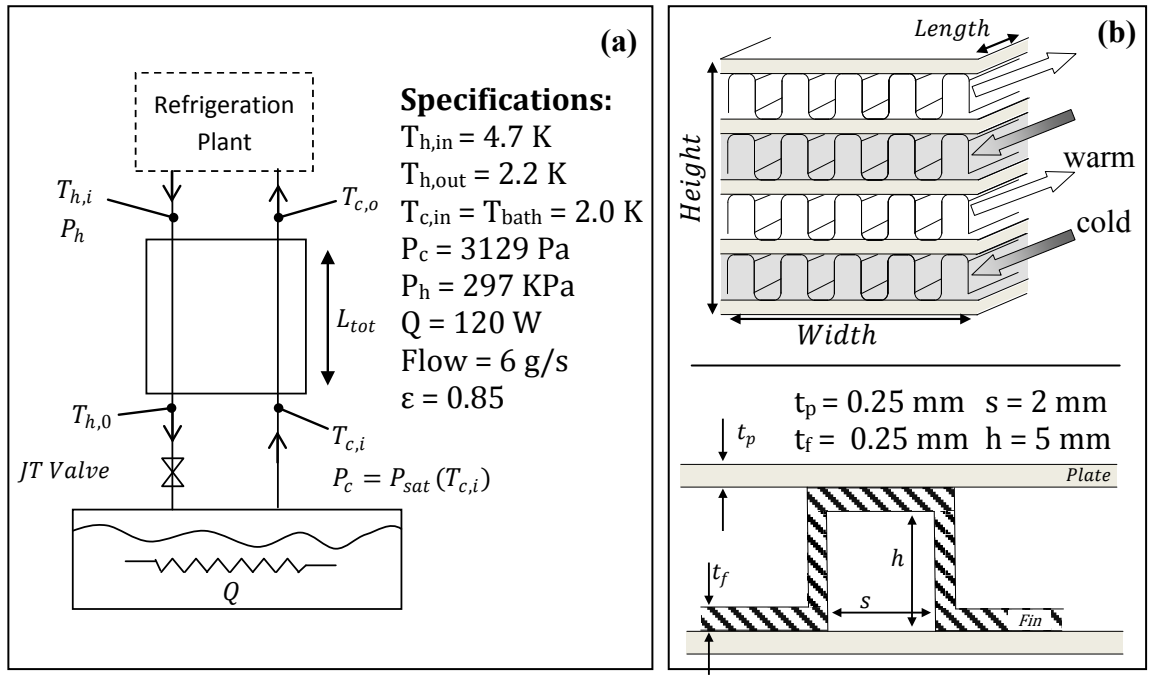

**FIGURE 1.** (a) Simplified flow schematic of plate-fin heat exchanger and example specifications.
(b) Illustration of PFHE rectangular duct geometry.

The supercritical helium then undergoes a final Joule-Thomson (J-T) expansion before entering the bath. The liquid fraction at the outlet of the J-T valve is related to the inlet temperature, where liquid fraction increases with a decrease in inlet temperature. Therefore, it is critical to use a high-effectiveness heat exchanger upstream of the J-T valve to maximize liquid fraction. Compactness is also important due to space limitations in the cryomodules and/or refrigeration plant.

Plate-fin heat exchangers have approximately ten times the surface to volume ratio compared to conventional tube-shell heat exchangers [1]. In cryogenic applications, axial conduction through the separating surfaces of the two streams can result in severe performance deterioration. Axial conduction acts to raise the average wall temperature and thus reduces the overall heat exchanger effectiveness [1-2]. The compactness of plate-fin heat exchangers results in relatively short conduction lengths between the warm and cold ends of the heat exchanger. Therefore, the design of compact and high-effectiveness cryogenic heat exchangers requires careful consideration of axial heat conduction. In addition, cryogenic fluid properties and metal conductivity change considerably with temperature and can lead to significant errors in models that assume constant properties. Herein, we present a numerical model that explicitly accounts for the effect of axial conduction and variable properties and discuss the initial results.

## NUMERICAL HEAT EXCHANGER MODEL

The model is based upon a numerical model described in detail by Nellis [2]. To explicitly model the effects of axial conduction and variable fluid properties, the heat exchanger is divided into several sections and the governing equations are derived in finite difference form. The location of the nodes and elements making up the computational grid of the heat exchanger model is illustrated in FIGURE 2.

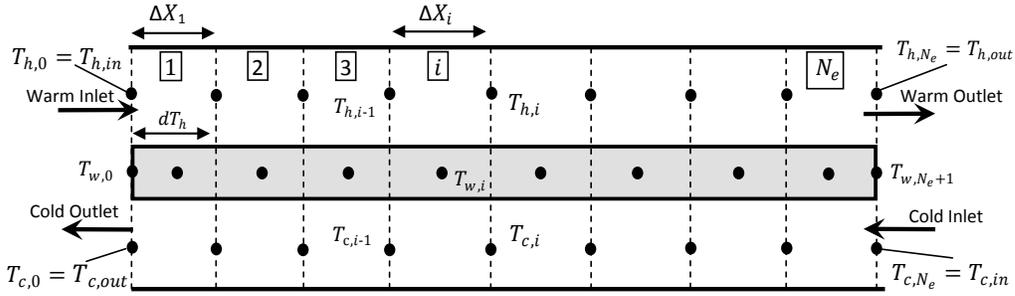

**FIGURE 2.** Computational grid used in the numerical model.

Knowing the supercritical helium inlet temperature supplied by the refrigerator and the specified warm side outlet temperature (i.e. effectiveness), the heat exchanger is divided into $N_e$ equal sections of temperature. Each section has three elements: warm fluid side, cold fluid side and the metal wall boundary. The temperature of the nodes for each section is assumed to be uniform in the lateral direction, thus simplifying to a quasi two-dimensional problem.

A typical set of heat exchanger specifications is presented in FIGURE 1a. The unknowns include: $N_e$ cold side temperatures ($T_{c,i}$), $N_e$ differential lengths ($\Delta X_i$) and $N_e+2$ metal wall temperatures, giving a total of $3N_e+2$ unknowns. To solve for the unknowns, control volumes are drawn around each of the three types of elements and a system of $3N_e$ energy balance equations are written and two additional equations are obtained by assuming adiabatic ends. The system of equations is written and iteratively solved using Engineering Equation Solver (EES) and a set of reasonable guesses [3]. The local helium fluid properties are obtained at each node using HePaK and look up tables were created for various metals at cryogenic temperatures [4].

## RESULTS

### Heat Exchanger Parameters

As an initial study, ordinary rectangular duct fin geometry was analyzed and the appropriate correlations from *Hesselgreaves* were used in the numerical model [1]. The heat exchanger specifications in FIGURE 1a were used as an example for this study. For a specified fin geometry, heat exchanger core width and height, and metal thermal conductivity, the heat exchanger length required to meet the specifications is calculated. The model has the ability to include parasitic heat loads and future studies will investigate the effects of including/excluding radiation shielding around the heat exchanger.

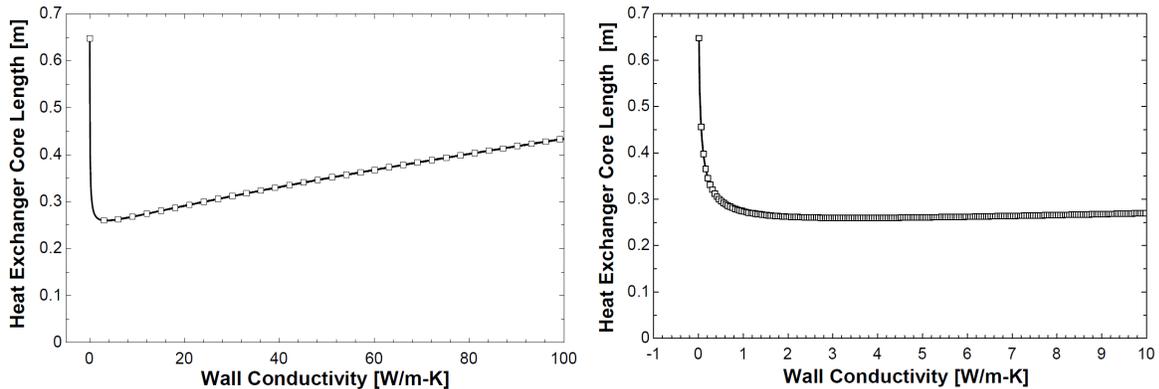

**FIGURE 3.** PFHE core length versus average wall conductivity graphed in two scales and using the specifications listed in FIGURE 1 ($\varepsilon=0.85$).

The results obtained in FIGURES 3 through 5 are based upon the specifications and fin geometry in FIGURE 1. A square heat exchanger core cross section of 0.3m x 0.3m was also used. The strong influence of axial conduction on heat exchanger design is illustrated in FIGURE 3, where the required heat exchanger core length versus an average thermal conductivity of the metal was calculated. At very low thermal conductivities, heat exchange between the hot and cold streams is hindered, thus decreasing effectiveness and requiring a longer heat exchanger to meet specifications. As the conductivity increases the required length decreases until it reaches a minimum of ~0.26 m at 3.5 W/m-K.

For thermal conductivities above 3.5 W/m-K, the effect of axial conduction begins to dominate and the required heat exchanger length again increases. While the optimal conductivity is around 3.5 W/m-K, zooming in on the minimum reveals that the conductivity can range from ~0.3-10 W/m-K without a large loss in performance.

The temperature profiles calculated for heat exchangers sized using variable conductivities of stainless steel (SS) type 304L and copper with a residual resistantance ratio of 120 (Cu120) are presented in FIGURE 4a and 4b [5-6]. The required heat exchanger core length is 0.83 m and 0.36 m for cores made of Cu120 and SS 304L, respectively. The high thermal conductivity of Cu120 reduces the overall temperature difference available within the heat exchanger and degrades the performance. Perhaps counter intuitively, the heat exchanger made of SS 304L performs better than one made from copper with the same dimensions. To further optimize the design, a metal with a thermal conductivity closer to the minimum in FIGURE 3a could be chosen. For example aluminum alloy 6061 is often used in cryogenic applications and has a thermal conductivity of ~3-5 W/m-K in the temperature range of interest [8]. FIGURE 4c is the temperature profile obtained for a heat exchanger sized using conductivity of Al 6061 [5]. By optimizing the thermal conductivity, the required length was further reduced to 0.26 m.

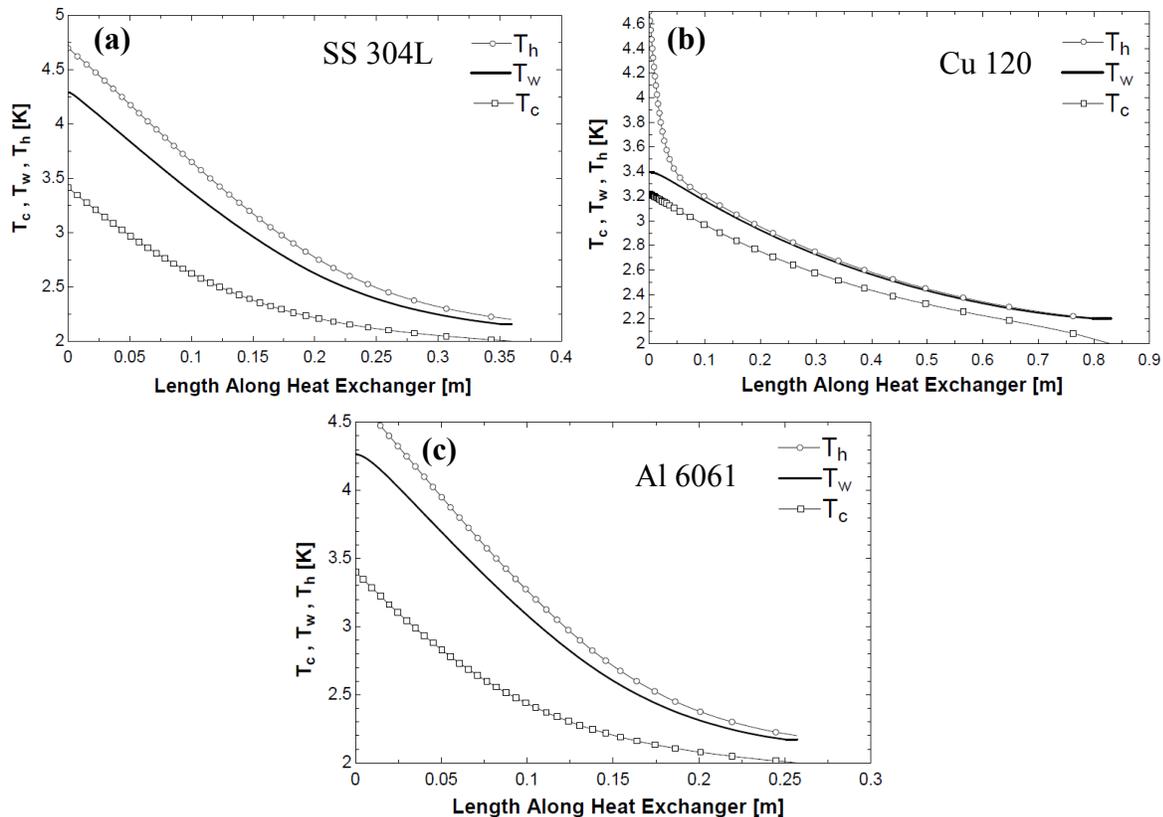

**FIGURE 4.** The calculated temperature profiles for a PFHE using the conductivities of (a) of SS 304L (b) Cu120 and (c) Al6061. All plots used the specifications and fin geometry listed in FIGURE 1 ($\varepsilon=0.85$).

The thermal conductivity of metals and dependence on temperature is largely influenced by the types of impurities, density and types of defects, percent cold working, annealing and manufacturing processes [7]. This is especially true for pure metals such as copper, where conductivities can vary by several orders of magnitude. For alloys, such as SS 304L and Al 6061, the variation in thermal conductivity from sample to sample is much smaller [8]. In addition to thermal conductivity, material cost and workability would be among deciding factors in final material selection.

Using a heat exchanger core cross section of 0.3m x 0.3m and the fin and plate thicknesses as specified in FIGURE 1, the required core length versus fin channel width was calculated at different channel heights. The results using the thermal conductivity of SS 304L are presented in FIGURE 5a. A decrease in channel width increases the number of fins and the total heat transfer surface area and therefore the required length decreases. For comparison, the analysis was repeated using the thermal conductivity of Cu120 and is presented in FIGURE 5b. Here the behavior is highly nonlinear. Small channel sizes increase the total number of fins and therefore the cross sectional area available for axial conduction. For very small channel sizes, the effect of axial conduction begins to dominate and the required heat exchanger length increases as channel size decreases.

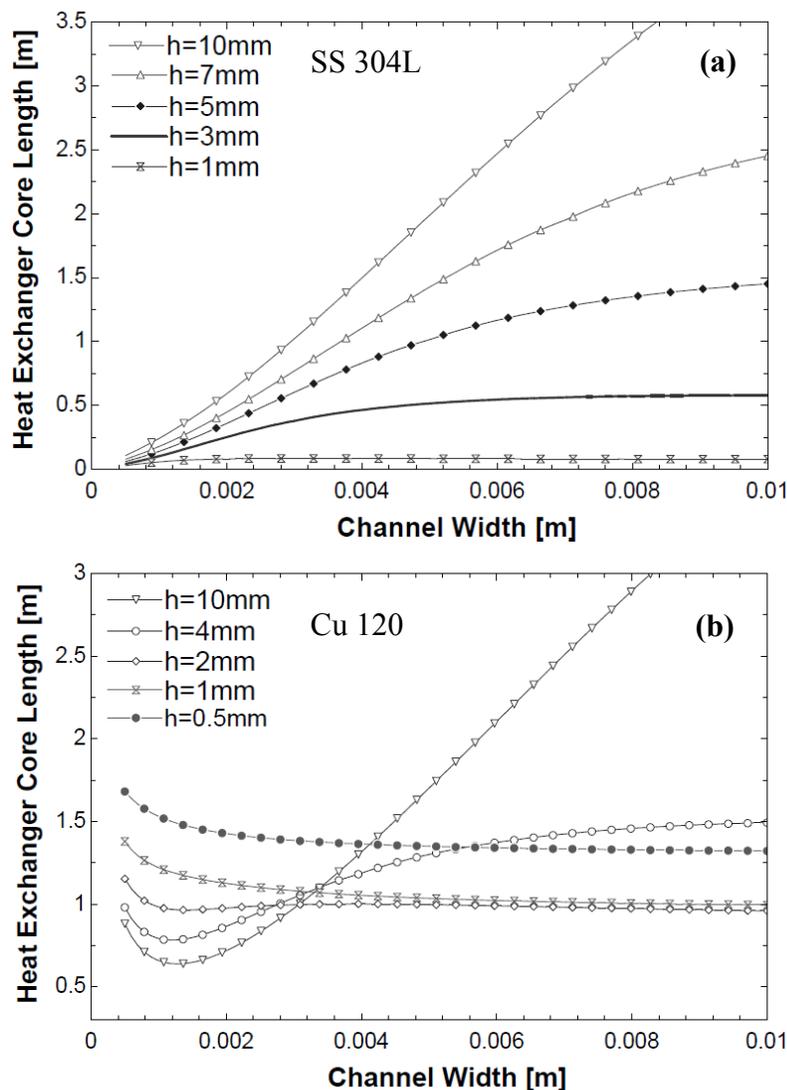

**FIGURE 5.** Heat exchanger core length versus the fin channel width at different channel heights calculated for (a) stainless steel and (b) copper using specifications listed in FIGURE 1 ($\varepsilon=0.85$).

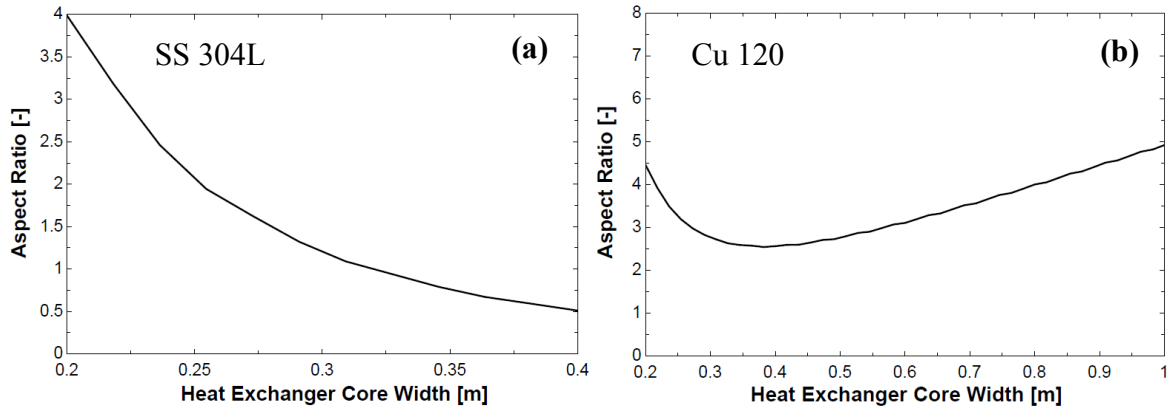
**FIGURE 6.** Heat exchanger aspect ratio versus core width calculated for (a) stainless steel and (b) copper using specifications in FIGURE 1 ($\varepsilon=0.85$).

Lastly, using the fin and channel geometry specified in FIGURE 1b and a square core cross section, the required heat exchanger core length versus core width was calculated. The results are plotted as aspect ratio (length over width) versus core width (FIGURE 6a-b). For a SS heat exchanger, the aspect ratio continuously decreases as core width decreases. An aspect ratio much less than one would typically be avoided, since an effective distributor design would be difficult to realize for small aspect ratios. For Cu120, the aspect ratio is much larger with a minimum of ~2.5 at a core width of 0.38m. As core width increases, the effect of axial conduction dominates due to an increase in available cross sectional area for axial conduction.

**Model Incorporated with J-T Expansion Process**

The J-T expansion process was modeled as an ideal isenthalpic expansion and was included in the overall heat exchanger model. The model was then used to analyze how liquid fraction obtained from J-T expansion varies with cold side inlet temperature.

For this analysis, a heat exchanger was first sized using the specifications in FIGURE 1 and the thermal conductivity of SS 304L. Next, using the fixed heat exchanger size, the model was adjusted to calculate the warm side outlet temperature and corresponding enthalpy. The liquid fraction was then determined using the calculated enthalpy and the superfluid bath pressure. The results are graphed as liquid fraction versus cold side inlet temperature (FIGURE 7a). As the cold side inlet temperature increases, the resulting liquid fraction decreases linearly. For cold inlet temperatures increasing from 2K to 2.2 K, the liquid fraction decreased from 0.856 to 0.839. This corresponds to a 0.12 g/s increase in supercritical helium supply required to maintain liquid level in the cooling bath.

The effect of variable heat load is also of interest. Using the heat exchanger sized for 120W, the liquid fraction and heat exchanger effectiveness were calculated as a function of heat load. The results are shown in FIGURE 7b. With heat loads ranging from 60 W to 300 W, the effectiveness ranged from ~0.92 to 0.64 and the liquid fraction from 0.88 to 0.77. Depending on the uncertainty in required heat load, the heat exchanger should be oversized to obtain a conservative design.

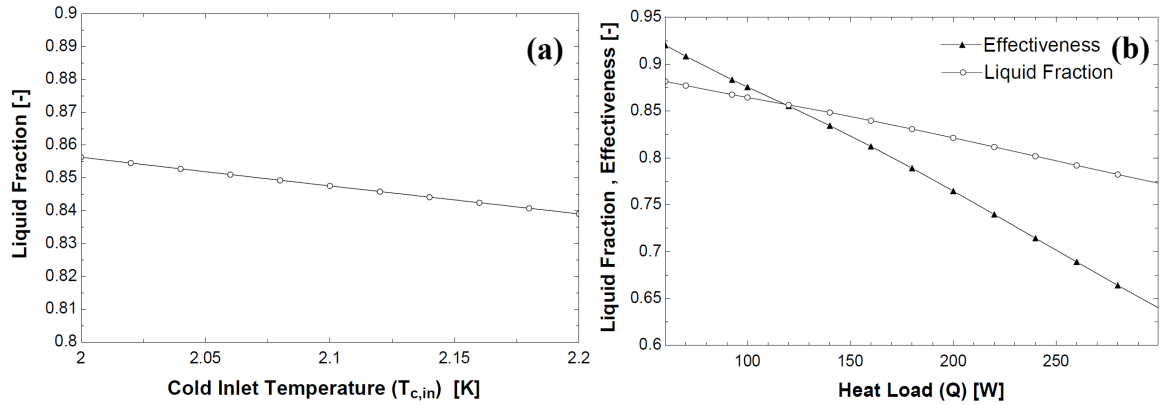

**FIGURE 7.** (a) Liquid fraction out of JT valve versus heat exchanger cold inlet temperature. (b) Liquid fraction out of JT valve and heat exchanger effectiveness versus heat load.

## FUTURE WORK

Location and size of the J-T heat exchanger are important factors that need consideration. Currently, two schemes are being considered: one that uses a large heat exchanger located at the plant, and the other that uses several smaller heat exchangers located at each cryomodule segment. In either scenario, the J-T valve will be located at the cryomodule, just before the inlet to the superfluid cooling bath. A detailed analysis will need to be carried out in order to determine the optimal location in terms of both capital and operational cost.

## SUMMARY

A numerical model that includes the effects of axial conduction and variable properties for a plate fin heat exchanger was developed and the effect of various design parameters on overall heat exchanger size was investigated. It was found that highly conductive metals should be avoided in the design of compact JT heat exchangers. For the geometry considered, the optimal conductivity is around 3.5 W/m-K and can range from 0.3-10 W/m-K without a large loss in performance. The model was implemented with an isenthalpic expansion process. Increasing the cold side inlet temperature from 2K to 2.2 K decreased the liquid fraction from 0.856 to 0.839 which corresponds to a 0.12 g/s increase in supercritical helium supply needed to maintain liquid level in the cooling bath. Lastly, it was found that the effectiveness increased when the heat load was below the design value. Therefore, the heat exchanger should be sized on the high end of the required heat load.

## ACKNOWLEDGEMENTS

The authors would like to thank G. Addison Merchut for his help with the graphics. Operated by Fermi Research Alliance, LLC under Contract No. DE-AC02-07CH11359 with the United States Department of Energy.